\documentclass[twoside]{article}
\usepackage{fleqn,espcrc2,epsf}
\usepackage{graphicx}
\usepackage{epsfig}
\usepackage[figuresright]{rotating}

\newcommand{\be}[1]{\begin{equation} \label{(#1)}}
\newcommand{\ee}{\end{equation}}
\newcommand{\ba}[1]{\begin{eqnarray} \label{(#1)}}
\newcommand{\ea}{\end{eqnarray}}

\newcommand{\AmS}{{\protect\the\textfont2
  A\kern-.1667em\lower.5ex\hbox{M}\kern-.125emS}}

\def\be{\begin{equation}}
\def\ee{\end{equation}}
\def\bea{\begin{eqnarray}}
\def\eea{\end{eqnarray}}
\def \KK {H.V.~Klapdor-Kleingrothaus}
\title{GENIUS - A New Underground Observatory for \hspace{5.cm}
	Non-Accelerator Particle Physics}

\author{H.V. Klapdor--Kleingrothaus
\address{Max--Planck--Institut 
f\"ur Kernphysik, P.O.Box 10 39 80, \\
D--69029 Heidelberg, Germany}\thanks{Spokesman of HEIDELBERG-MOSCOW and GENIUS Collaborations,  
	e-mail:klapdor@gustav.mpi-hd.mpg.de, 
	home page: http://mpi-hd.mpg.de.non\_acc/}
}

\begin{document}

\begin{abstract}
	The GENIUS (\underline {Ge}rmanium in Liquid 
	\underline {Ni}trogen \underline {U}nderground \underline {S}etup) 
	project has been proposed in 1997 
\cite{KK-BEY97} 
	as first third generation double beta decay project, with 
	a sensitivity aiming down to a level of an effective neutrino 
	mass of 
$\langle m_{\nu} \rangle <$ 0.01\,eV or less. 
	Such sensitivity is important to fix the structure of the 
	neutrino mass matrix with high accuracy, 
	which cannot be done by neutrino oscillation 
	experiments alone.
	GENIUS will allow broad access also to many other topics of 
	physics beyond the Standard Model of particle physics at the 
	multi-TeV scale. For search of cold dark matter GENIUS will cover 
	a large part of the parameter space of predictions of SUSY 
	for neutralinos as dark matter 
\cite{Bed-KK2,Ell,KK-LowNu2}. 
	Finally, GENIUS has the potential to be a real-time detector 
	for low-energy (pp and $^7{Be}$) solar neutrinos 
\cite{Bau-KK,KK-LowNu2}. 
	A GENIUS-Test Facility has just been funded and will come into 
	operation by end of 2002.
\vspace{1pc}
\end{abstract}

\maketitle


\section{INTRODUCTION}

	Without double beta decay there can be 
	no solution of the nature of the neutrino (Dirac or Majorana 
	particle) and of the structure of the neutrino mass matrix. Only 
	investigation of $\nu$ oscillations {\it and} double beta decay 
	together can lead to an absolute mass scale.

	Concerning the search for cold dark matter, even a discovery of 
	SUSY by LHC will not have proven that neutralinos form 
	indeed the cold dark matter in the Universe. Direct detection 
	of the latter by underground detectors remains indispensable. 
	Concerning solar neutrino physics, present information on possible 
	$\nu$ oscillations relies on 0.2$\%$ 
	of the solar neutrino flux. The total pp neutrino flux has not 
	been measured and also no real-time information is available for 
	the latter.

	The GENIUS project proposed in 1997 (see 
\cite{KK-BEY97,KK60Y})
	as the first third generation $\beta\beta$ detector, could attack 
	all of these problems with an unpredented sensitivity.


\section{GENIUS, DOUBLE BETA DECAY \\
	AND THE LIGHT MAJORANA \\
	NEUTRINO MASS}

\begin{figure}[htb]
%
\centering{
\includegraphics*[scale=0.27]
{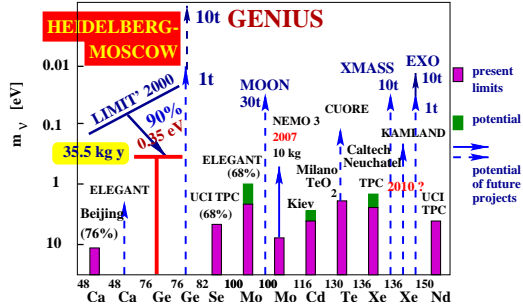}
}

\vspace{-0.7cm}
\caption{Present situation, and expectation for the future, of the 
	most promising $\beta\beta$ experiments. For the Heidelberg-Moscow 
	experiment - old result (2000) 
\cite{HDM01}. 
	Framed parts of the bars: 
	present status; open parts: expectation for running experiments; 
	solid and dashed lines: experiments under construction  
	or proposed experiments. For references see 
\cite{KK60Y,KK-NANP}.
\label{Status}}
\end{figure}

\begin{figure}[htb]
\vspace{9pt}
\centering{
\includegraphics*[width=75mm, height=40mm]
{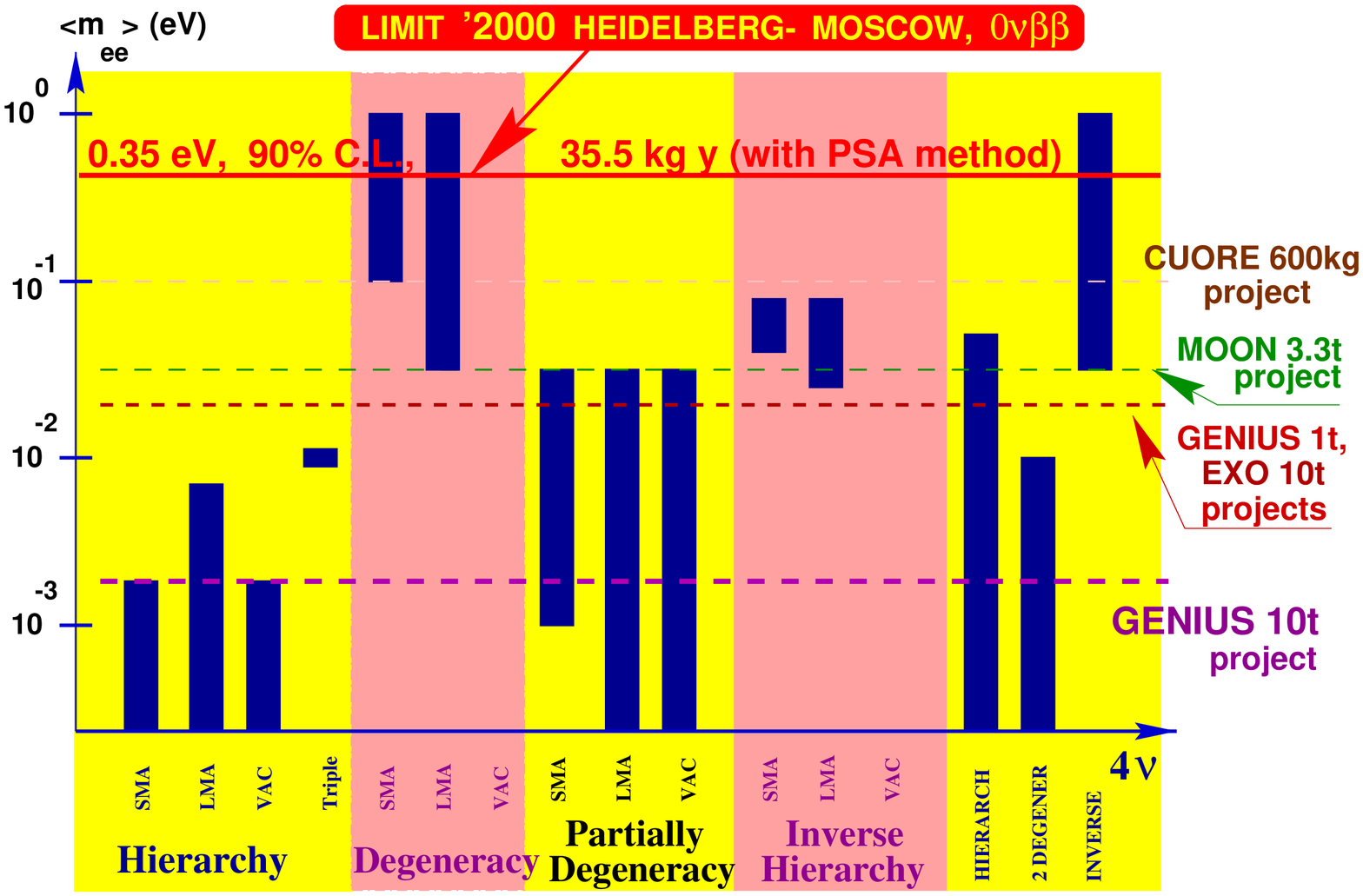}
}

\vspace{-0.5cm}
\caption{Values expected from $\nu$ oscillation experiments for 
	$\langle m \rangle$ 
	in different schemes. 
	The expectations are compared with the neutrino mass limits 
	{\it obtained} from the 
	HEIDELBERG-MOSCOW experiment as well as the {\it expected}  
	sensitivities for the CUORE, MOON, EXO proposals and the 1 ton 
	and 10 ton proposal of GENIUS 
\cite{KK-BEY97}. 
	For references  and more details about the different experiments see 
\cite{KK60Y,KK-NANP}.
\label{Sum-difSchemNeutr}}
\end{figure}

	Among present double beta experiments the 
	the most sensitive experiment is {\it since eight years} the 
	HEIDELBERG-MOSCOW experiment using the world's largest source 
	strength of 11\,kg of 86$\%$ enriched $^{76}{Ge}$ 
	in form of 5 high-purity Ge detectors, run in the Gran Sasso 
	Underground laboratory.
	This experiment yields after 53.9 (35.5)\,kg\,y 
	of measurement a half-life limit of 
	${\rm T}_{1/2}^{0\nu} = 1.3\times 10^{25}(1.9\times 10^{25})$
	${\rm y}$ ,(90$\%$ c.l.) 
	and an upper limit for the effective neutrino mass 
	$\langle m \rangle $ of 0.42(0.35)\,eV 
\cite{HDM01}. 
	The numbers in parentheses are deduced from PSA. 
	These numbers are just entering into the range 
	of expectations for $\langle m \rangle$ 
	from neutrino oscillation experiments 
(see Fig.\ref{Sum-difSchemNeutr}).
	New approaches and considerably enlarged experiments 
	(as discussed, e.g. in 
\cite{KK-NANP}) 
	are required, however, to improve the present accuracy. 
	This can not be done by any of the presently operated 
	double beta experiments, whose status is shown in 
Fig.\ref{Status}, 
	together with the potential of some 
	future projects under discussion.
	
	With the era of the HEIDELBERG-MOSCOW experiment the time 
	of the small smart experiments is over.

	Table\,\ref{Key-Exper} 
	lists some key numbers for GENIUS, and of the main other proposals 
	made after the GENIUS proposal. 
	Not all of these proposals fully cover the region 
	to be probed. 
		Since it was realized in the HEIDELBERG-MOSCOW experiment, 
	that the remaining small background is coming from the material 
	close to the detector (holder, copper cap, ...), 
	elimination of {\it any} material close to the detector 
	will be decisive. Experiments which do not take this 
	into account, like, e.g. CUORE 
	and MAJORANA, 
	will allow only rather limited steps in sensitivity.

	Another crucial point is the energy resolution, 
	which can be optimized {\it only} in experiments 
	using Germanium detectors or bolometers. 
	It will be difficult to probe evidence for this rare decay 
	mode in experiments, which have to work - as result of their 
	limited resolution - with energy windows around 
	Q$_{\beta\beta}$ of several hundred keV, such as NEMO III, 
	EXO. 

	In the first proposal for a third generation double 
	beta experiment, the GENIUS proposal 
\cite{KK-BEY97},
	the idea is to use 'naked' Germanium detectors in a huge tank 
	of liquid nitrogen. 
	It has been shown that the detectors show 
	excellent performance under such conditions
\cite{KK-BEY97}.
	GENIUS seems to be at present the {\it only} 
	proposal, which can fulfill {\it both} requirements mentioned above. 
	The potential of GENIUS is together with that of some later proposals 
	indicated in Fig. 
\ref{Sum-difSchemNeutr}. 
	
	For technical questions and extensive Monte Carlo simulations of the 
	GENIUS project for its various applications 
	we refer to 
\cite{KK-BEY97}.


\begin{figure}[h]
\begin{center}
\epsfig{file=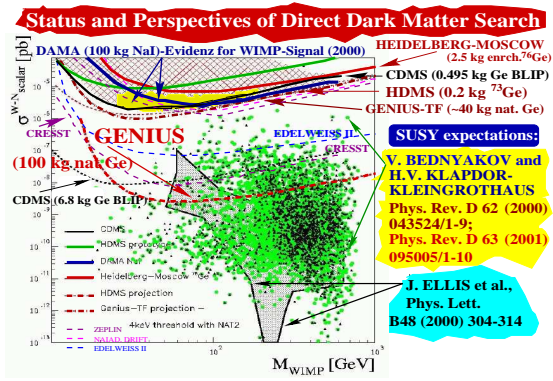,width=73mm, height=50mm}
\end{center}

\vspace{-0.5cm}
\caption[]{
       WIMP-nucleon cross section limits in pb for scalar interactions as 
       function of the WIMP mass in GeV. 
       Shown are contour lines of present experimental limits (solid lines) 
       and of projected experiments (dashed lines). 
       Also shown is the region of evidence published by DAMA. 
       The theoretical expectations are shown 
	for the MSSM, by two scatter plots, 
	- for accelerating and for non-accelerating Universe (from  
\cite{Bed-KK2}), 
	and for MSUGRA, by the grey region (from  
\cite{Ell}).
	{\em Only}~ GENIUS will be able to probe the shown range 
       also by the signature from seasonal modulations.
\label{fig:Bedn-Wp2000}}
\end{figure}


\begin{table*}[t]
\caption{\label{Key-Exper}
	Some key numbers of future double beta decay experiments (and of 
	the {\sf HEIDELBERG-MOSCOW} experiment). Explanations: 
	${\nabla}$ - assuming the background of the present pilot project. 
	$\ast\ast$ - with matrix element from [Sta90*-II], [Tom91**-I], 
	[Hax84**-I], [Wu91*-II], [Wu92*-II] (see Table II in [HM99*-III]), 
	and allowing for an uncertainty of $\pm$ 50$\%$ 
	of these matrix elements. 
	${\triangle}$ - this case shown 
	to demonstrate {\bf the ultimate limit} of such experiments. 
	For details see \cite{KK60Y}.}
\newcommand{\m}{\hphantom{$-$}}
\newcommand{\cc}[1]{\multicolumn{1}{c}{#1}}
\renewcommand{\tabcolsep}{0.42pc} 
\renewcommand{\arraystretch}{.1} 
{\footnotesize
\m
{ 
\begin{center} 
\begin{tabular}[!h]{|c|c|c|c|c|c|c|c|}
\hline
 &  &  &  & Assumed backgr.  & Run- & 	Limits & 	Limits \\
$\beta\beta$-- & & & Mass & $\dag$ events/kg y keV,  & ning 
&	for $0\nu\beta\beta$	 & 	for\\
$Isoto-$ & $Name$ & $Status$ & $(ton-$ & $\ddag$ ev./kg y FWHM & Time  
& 	half-life (y) & 	 $\langle m_{\nu} \rangle$(eV) \\
pe & & & nes) &  $\ast$ events/yFWHM & (tonn. & 
& 
\\ 
& & & & 	& years) &  	& \\
\hline
 &  &  &  &  &  &  & \\
~${\bf ^{76}{Ge}}$ & {\bf HEIDELBERG $^1$} & {\bf run-}  & 0.011 & $\dag$ 0.06 
& {\bf 53.9} & ${\bf 1.3x~{10}^{25}\,y}$ & {\bf 0.42\,eV}\\
 & {\bf MOSCOW}  &  &  (enri-  &  &  {\bf kg y} 
& ${\bf 1.9x~{10}^{25}\,y}$,PSA  
& {\bf 0.35\,eV},PSA\\
& {\bf \cite{KK-Evid,HDM01,KK60Y}} & {\bf ning} & ched) 
& $\ddag$ 0.24~~~~$\ast$ 2   &  
& {\bf 90$\%$ c.l.}  & {\bf 90$\%$ c.l.}\\
\hline
\hline
 &  &  &  &  &  &  & \\
${\bf ^{100}{Mo}}$ & {\sf NEMO III} & {\it under} & $\sim$0.01 & $\dag$ 
{\bf 0.0005} &  &  &\\
 & {\tt [NEM2000]}& {\it constr.} & (enri- & $\ddag$ 0.2  & 50 & 
${10}^{24-25}$ & 0.3-0.7\\
 &  &  & -ched) &  $\ast$ 2 &kg y  &  &\\
\hline
&  &  &  &  &  &  & \\
${\bf ^{130}{Te}}$ & ${\sf CUORE}^{\nabla}$ & Pro- & 0.75 & $\dag$ 0.5 & 5 & 
$9\cdot{10}^{24}$ & 0.2-0.5\\
 & {\tt [Gui99a*-VI]}& posal &(nat.)  & $\ddag$ 4.5~~~~ $\ast$ 1000 &  & & \\ 
\hline
&  &  &  &  &  &  &  \\
${\bf ^{130}{Te}}$ & {\sf CUORE}  &  Pro- & 0.75   & $\dag$ 0.005 & 5 
& $9\cdot{10}^{25}$ & 0.07-0.2\\
&  &  posal  & (nat.)  &  $\ddag$ 0.045~~~~$\ast$ 45 &  & &\\
\hline
&  &  &  &  &  &  & \\
${\bf ^{100}{Mo}}$ & {\sf MOON} & idea & 10 (enr-& ? & 30 & ? & 0.03\\
 & {\tt [Eji99b*-VI]} &  & 100(nat.) &  & 300 & & \\
\hline
&  &  &  &  &  &  & \\
${\bf ^{136}{Xe}}$ & {\sf EXO} & Pro-& 1 & $\ast$ 0.4 & 5 & 
$8.3\cdot{10}^{26}$ & 0.05-0.14\\
&  & &  &  &  &  & \\
${\bf ^{136}{Xe}}$ & {\tt [Dan2000a]} & posal  & 10 & $\ast$ 0.6 & 10 & 
$1.3\cdot{10}^{28}$ & 0.01-0.04\\
&  &  &  &  &  &  & \\ 
\hline 
\hline
&  &  &  &  &  &  &  \\
~${\bf ^{76}{Ge}}$ & {\bf GENIUS} & Pro- & 1  & $\dag$ 
${\bf 0.04\cdot{10}^{-3}}$ & 1 & ${\bf 5.8\cdot{10}^{27}}$ & 
{\bf 0.02-0.05} \\
 & {\tt \cite{KK-BEY97}}  & posal &(enrich.)  
& $\ddag$ ${\bf 0.15\cdot{10}^{-3}}$ & & & \\
&  &  &  & $\ast$ {\bf 0.15} &  &  &  \\
&  &  & 1 & ${\bf \ast~ 1.5}$ & 10 & ${\bf 2\cdot{10}^{28}}$  & 
{\bf 0.01-0.028} \\
\hline
&  &  &  &  &  &  &  \\
~${\bf ^{76}{Ge}}$ & {\bf GENIUS} & Pro- & 10 
& $\ddag$ ${\bf 0.15\cdot{10}^{-3}}$ & 10 &
${\bf 6\cdot{10}^{28}}$ & {\bf 0.006-0.016}\\
&  {\tt \cite{KK-BEY97}} &  posal & (enrich.) &  ${\bf 0^{\triangle}}$	
&  10  &  ${\bf 5.7\cdot{10}^{29}}$	&  {\bf 0.002-0.0056}\\
\hline 
\end{tabular}\\
\end{center} 
}}
\end{table*}



\vspace{-.3cm}
\section{GENIUS AND OTHER BEYOND STANDARD MODEL PHYSICS}

	GENIUS will also allow 
	the access to a broad range of other beyond 
	SM physics topics in the multi-TeV range. Already now 
	$\beta\beta$ decay probes the TeV scale on 
	which new physics should manifest itself (see, e.g. 
\cite{KK-BEY97,KK-TR98}). 
	Basing to a large extent on the theoretical work of the Heidelberg 
	group in the last six years, the HEIDELBERG-MOSCOW experiment yields 
	results for SUSY models (R-parity breaking, sneutrino mass), 
	leptoquarks (leptoquarks-Higgs coupling), compositeness, 
	right-handed W mass, nonconservation of Lorentz invariance and 
	equivalence 
	principle, mass of a heavy left or righthanded neutrino, 
	competitive to corresponding results from high-energy accelerators 
	like TEVATRON and HERA. The potential of GENIUS could extend into the 
	multi-TeV region for these fields and its sensitivity would 
	correspond to that of LHC or NLC and beyond (for details see 
\cite{KK60Y,KK-TR98}).


\vspace{-0.47cm}
\section{GENIUS AND\\ COLD DARK MATTER SEARCH}

	GENIUS would in a first step, with 100 kg of {\it natural} Ge 
	detectors, and in three years measurring time, 
	cover a significant part of the MSSM parameter space 
	for prediction of neutralinos as cold dark matter 
(Fig.\ref{fig:Bedn-Wp2000}). 
	For this purpose the background 
	in the ener\-gy range $<$\,100\,keV has to be reduced to 
	${10}^{-2}$\,events/\,kg\,y\,keV, which is possible if the detectors 
	are produced and handled on Earth surface under heavy shielding, 
	to reduce the cosmogenic background produced by spallation through 
	cosmic radiation to a minimum. For details we refer to 
\cite{KK-BEY97,KK-LowNu2}.
Fig.\ref{fig:Bedn-Wp2000}
	 shows together with the expected sensitivity of GENIUS predictions 
	for neutralinos as dark matter by two models, one basing on 
	supergravity
\cite{Ell}, 
	the other on the MSSM,
	with more relaxed unification conditions  
\cite{Bed-KK2}.

	The sensitivity of GENIUS for Dark Matter 
	with 100\,kg of natural Germanium 
	is better than that 
	obtainable with a 1\,${km}^3$ AMANDA detector for 
	{\it indirect} detection (neutrinos 
	from neutralino annihilation at the Sun). Interestingly both 
	experiments would probe different neutralino 
	compositions: GENIUS mainly gaugino-dominated neutralinos, 
	AMANDA mainly neutralinos with comparable 
	gaugino and Higgsino components. 
	It should be emphasized that, together with DAMA, 
	GENIUS will be the 
	{\it only} future Dark Matter experiment, 
	which would be able to positively 
	identify a dark matter signal by the seasonal modulation signature. 
	This {\it cannot} be achieved, for example, by the CDMS experiment.


\vspace{-.3cm}
\section{GENIUS AND LOW-ENERGY\\
	SOLAR NEUTRINOS}

	No experiment has 
	separately measured the pp and $^7{Be}$ neutrinos and no experiment 
	has measured the {\it full} pp $\nu$ flux. BOREXINO plans to 
	measure $^7{Be}$ neutrinos, the access to pp neutrinos 
	being limited by $^{14}C$ contamination (the usual problem of 
	organic scintillators). GENIUS could be the first detector measuring 
	the {\it full} pp ( and $^7{Be}$) neutrino flux in real time.

\begin{figure}[htb]
\vspace{-9pt}
\centering{
\includegraphics*[width=55mm, height=77mm, angle=-90]{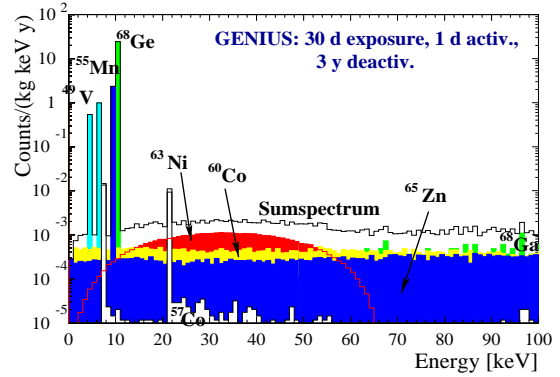}} 

\vspace{-0.7cm}
\caption{Simulated cosmogenic background during detector production. 
	Assumptions: 30\,days exposure of material before processing, 
	1\,d activation after zone refining, 3\,y deactivation underground 
	(see \cite{KK-LowNu2}).
\label{Backgr-Sol}}
\end{figure}

	With a radius of GENIUS of 13\,m and improving some of the 
	shielding parameters as described in 
\cite{Bau-KK,KK-LowNu2} 
	the background can be reduced to a level of 
${10}^{-3}$ events/\,kg\,y\,keV 
(Fig.\ref{Backgr-Sol}).  
	This will allow to look for the pp and $^7{Be}$ solar neutrinos by 
	elastic neutrino-electron scattering with a threshold\- 

\begin{figure}[htb]
\vspace{-0.5cm}
\centering{
\includegraphics*[width=70mm, height=45mm]{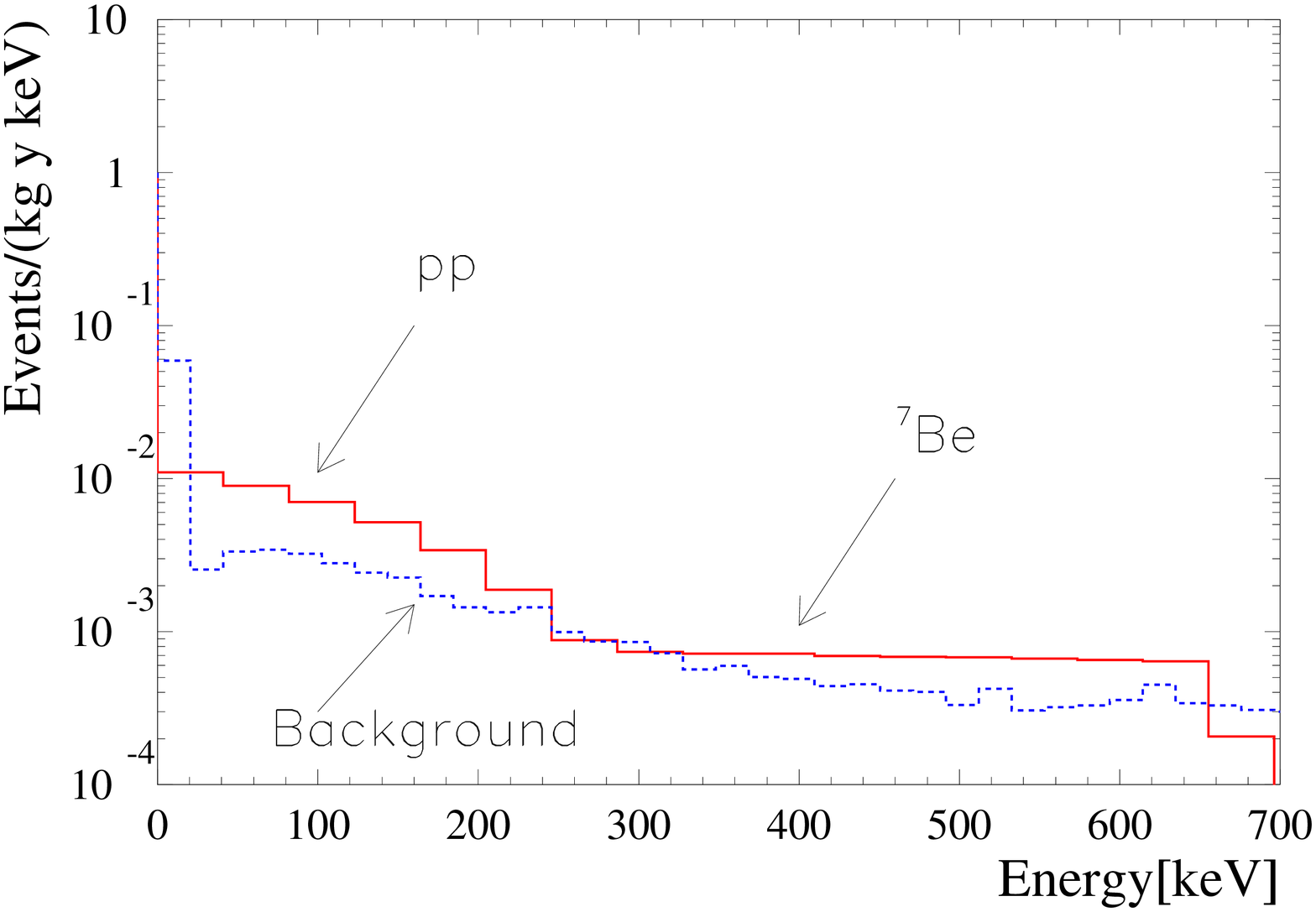}}

\vspace{-0.5cm}
\caption{Simulated spectrum of low-energy solar neutrinos (according to SSM) 
	for the GENIUS detector (1\,tonne of natural or enriched Ge) 
	(according to 
\cite{Bau-KK,KK-LowNu2}).
\label{Sol-Spectr}}
\end{figure}

\noindent
	of 11\,keV or at 
	most 19\,keV (limit of possible tritium background) 
(Fig.\ref{Sol-Spectr}), 
	which would be the 
	lowest threshold among other proposals to detect pp 
	neutrinos, such as HERON, HELLAZ, NEON, LENS, MOON, XMASS.

	The counting rate of GENIUS (10\,ton) would be 6 events per day 
	for pp and 18 per day for $^7{Be}$ neutrinos, i.e. similar to 
	BOREXINO, but by a factor of 30 to 60 larger than a 20\,ton LENS 
	detector and a factor of 10 larger than the MOON detector.


\vspace{-0.50cm}
\section{GENIUS - TEST FACILITY}

	Construction of a test facility for GENIUS - GENIUS-TF - 
	consisting of $\sim$ 40\,kg of HP Ge detectors suspended in a 
	liquid nitrogen box has been started. Up to middle of 2001, 
	six detectors each of $\sim$ 2.5\,kg 
	and with a threshold of as low as 
	$\sim$ 500\,eV have been produced.

\begin{figure}[htb]
\vspace{-0.5cm}
\centering{
\includegraphics*[width=75mm, height=51mm]{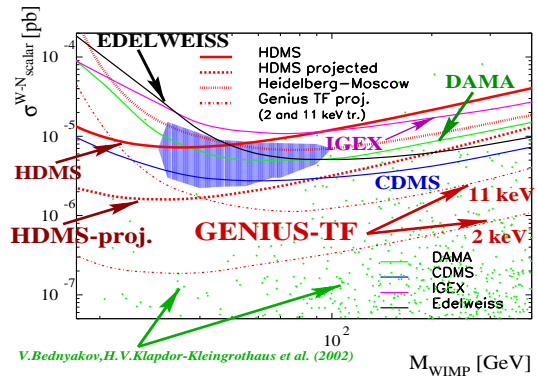}}

\vspace{-0.6cm}
\caption{The potential of GENIUS-TF for dark matter search. 
	  WIMP-nucleon cross section limits as a function of the WIMP
  	mass for spin-independent interactions. 
  	The solid lines are current limits 
	of the HEIDELBERG-MOSCOW experiment, 
  	the HDMS (Heidelberg Dark Matter Search), 
	the DAMA  
	and the CDMS experiments. 
  	The dashed curves are the expectation for HDMS, 
	and for Genius-TF
  	with an energy threshold of 11\,keV and 2\,keV (no tritium
  	contamination), respectively, and a 
  	background index of 2\,events/kg\,y\,keV below 50\,keV.
  	The  filled contour represents the 
	evidence region of the DAMA experiment. 
\label{fig:Upper-Gen-TF}}
\end{figure}

	Besides test of various parameters of the GENIUS project, the test 
	facility would allow, with the projected background of 
	2-4\,events/\,kg\,y\,keV in the low-energy range, 
	to probe the DAMA evidence 
	for dark matter by the seasonal modulation signature (Fig. 
\ref{fig:Upper-Gen-TF}).
	For details we refer to
\cite{KK2000}. 


\vspace{-0.5cm}
\section{CONCLUSION}

	The GENIUS project is - among the projected or discussed third 
	generation double beta detectors - 
	the one which may exploit this method 
	to push the sensitivity on the neutrino mass to the ultimate limit. 

	GENIUS is the only one of the new projects which 
	simultaneously has a huge potential for cold dark matter search, 
	{\it and} for real-time detection of low-energy neutrinos. 


\end{document}